
\line{\hfill TAUP-1937-91}
\vskip .1in
\line{\hfill Dec,1991}
\date{}
\vskip 1 true cm
\titlepage
\title{\bf Exact Solutions of Four Dimensional
 Black Holes in String Theory}
\vskip 1 true cm
\centerline{\caps D.~Gershon}
\centerline{School of Physics and Astronomy}
\centerline{Beverly and Raymond Sackler Faculty of Exact Sciences}
\centerline{{Tel-Aviv University, Tel-Aviv 69978, Israel.}\footnote
{\dagger}{e-mail: GERSHON@TAUNIVM.BITNET}}
\vskip 2 true cm
\abstract
\bigskip
\noindent

 We construct an exact conformal field theory as an
 $SL(2,R)\times SU(2)/U(1)^2$ gauged
 WZW model,  which describes
 a black hole   in four dimensions.
 Another exact
 solution, describing a black membrane in four dimensions
 (in the sense that the event horizon is an infinite plane) is
 found as an $SL(2,R)\times U(1)^2/U(1)$ gauged WZW model.
 Both the black hole and the black membrane carry axionic charge.
 Finally, we construct an exact solution of a 4D black hole
 with electromagnetic field, as an
$SL(2,R)\times SU(2)\times U(1)/U(1)^2$ gauged WZW model.
This black hole carries both electric   and axionic charges.
\endpage
{\bf 1.  Introduction}

String theory is the first consistent quantum theory
 that accomodates            gravity. However,
it is not clear wheather  it could be established as the
quantum theory of gravity.

 As a theory of gravity we would expect that string theory would
 exhibit features encountered in general relativity.
Such an important characteristic would,
for example, require that string theory could
 describe objects auch as black holes.

Recently, it was shown by Witten [1] that string theory of two
dimensional target space can describe black holes in two dimensions.
This was obtained as a WZW action of the coset $SL(2,R)/U(1)$.
An exact solution of charged black holes was shown later
[2] as a WZW of
the coset $SL(2,R)\times U(1)/U(1)$)
However, in order to have a more realistic relation
between string theory and
nature, we would like to see that string theory can
describe black holes
in four dimensions. (In three dimensions an exact
solution of black
strings was obtained [3]).
  It was shown in [4]
 that the solution of the equations of motion
 of the effective
action in string theory in four dimensions has      a
solution of a
charged black hole.

In this work we  derive  string theories that have background
 of a black holes in four dimensions.
 We show the exact
conformal field theories that yield the 4D target space metric
which         describe black hole.
The first solution is a gauged WZW model of the
$SL(2,R)\times SU(2)/   U(1)^2$ coset theory.
 The black hole
found here is axisymmetric and carries  axionic
charge.

We further obtain a solution of a four dimensional black
membrane, by taking a WZW    $SL(2,R)\times U(1)^2/U(1)$ coset.
This solution describes a point singularity surrounded by an
infinite area event horizon.
Finally, we  give a solution of a black hole  with
 electromagnetic field by enlarging the coset
models. The charged black hole is obtained as an
$SL(2,R)\times U(1)\times SU(2)/U(1^2$ model (and the charged black
membrane as an $ SL(2,R)\times U(1)^3/U(1)$) and carries also axionic
charge.
\endpage
{\bf 2.  Derivation of 4D Black Hole Solution}

First, let us derive      the conformal theory which
deacribes the black
hole.
  We start with a Wess Zumino Witten action $\lbrack 5\rbrack$
of the group $SL(2,R)\otimes SU(2)$.
$$L_0(g)={k\over 2\pi}\int_{\Sigma}d^2\sigma
Tr(g_{-1}\partial_+ g g^{-1}
\partial_- g) -\Gamma  \eqno(1) $$   $g$ is an
element of a group
$G$ and $\Gamma$ is the Wess Zumino
term $$\Gamma={k\over 12\pi}\int_B Tr(g^{-1}dg\wedge
g^{-1}dg \wedge
g^{-1}dg) \eqno(2)$$ where B is the manifold whose boundary
 is $\Sigma$.
We take $h_1\in SL(2,R)$ and $h_2\in SU(2)$.
Since we have a direct product, the lagrangian is now
$L_0(g)=L_0(h_1)+L_0(h_2)$. For  compact groups $k$ should
be integer for topological reasons (second reffernce in
$\lbrack 5\rbrack$),
 therefore we take $k_1$
for
the non-compact $SL(2,R)$ group in $L_0(h_1)$
and an integer $k_2$ for the $SU(2)$ group in $L_0(h_2)$.
We shall now derive the black hole solution in Lorenzian
 signature.
   We  parametrise the $SL(2,R)$ group
by $$h_1=\left(
\matrix {a&u\cr -v&b\cr}\right)\eqno(3)$$
with $ab+uv=1$  and we parametrize $SU(2)$ as
 $$h_2=\exp(i{\phi_L\over 2}\sigma_3)\exp(i\theta
 \sigma_1)\exp(i{\phi_R\over 2}\sigma_3)\eqno(4)$$
  First we gauge a $U(1)$ subgroup of $SU(2)$ generated by
 $i\sigma_3$. Let as first treat only the
  part containing the $SU(2)$ group.
The gauged WZW action is $\lbrack 6\rbrack$
$$L(A,h_2)=L_{0,k_2}(h_1)+{k_2
\over 2\pi}\int_{\Sigma} d^2\sigma Tr(A_+
\partial _-h_2 h_2^{-1}+A_-h_2^{-1}\partial_+h_2 $$ $$+
A_+A_- +A_+h_2A_- h_2^{-1})
 \eqno(6)$$
(where $A_i$ is the gauge field that gauges the symmetry
$h_2\rightarrow h h_2 h^{-1}$, with $h$ being
 generated by $i\sigma_3$.)
 The gauged action is
$$L_2(A_i,\theta,\phi_L,\phi_R)=S_{WZW}(\theta,\phi_L,\phi_R)+
 {k_2\over 2\pi}\int d^2\sigma \{A_-(\partial_+\phi_R+\cos 2\theta
 \partial_+\phi_L)$$
$$+A_+(\partial_-\phi_L+\cos 2\theta\partial_-\phi_R
)-4A_-A_+(\cos ^2\theta)\}
 \eqno(6)$$
with $$S_{WZW}( \theta,\phi_L,\phi_R)=
 {k_2\over 8\pi}\int d^2\sigma(4\partial_-\theta\partial_+\theta
 +\partial_-\phi_L\partial_+\phi_L +\partial_-\phi_R\partial_+
 \phi_R$$ $$
 +2\cos 2\theta \partial_-\phi_L\partial_+\phi_R)\eqno(7)$$
In this action we can fix a gauge $\phi_L=\phi_R=\phi$.
Then we integrate the gauge fields out
and obtain to order one loop in $k$  $\lbrack 7\rbrack$
$$S=\int D\lbrack \theta,\phi \rbrack e^{L_{g.f.}(h_2)}\det
\lbrack {-2\pi^2\over k_2\cos^2\theta}\rbrack\eqno(8)$$
where
$$L_{g.f.}(h_2)= {k_2\over 2\pi}\int d^2\sigma
(\partial_+\theta\partial_-\theta+\cos^2\theta\partial_+\phi
\partial_-\phi)\eqno(9)$$
We shall treat the determinant in the end of the derivation.
 It is important, however,
that it is independent of $\phi$.

Now we return to the full action.
 In the following we obtain the black hole solution in Lorenzian
 signature.
We gauge a $U(1)$ subgroup of $SL(2,R)$ generated by
 $\left (\matrix {1&0\cr 0&-1\cr}\right)$ together with a
           translational gauge of $\phi$ in  (9)
 (the gauged action  exhibits a translational
symmetry in $\phi$).
 To be explicit, we gauge the following symmetry transformation:
$$\delta a=2\epsilon a, \delta b =-2\epsilon b,
\delta u=\delta v=0,
 \delta \phi=2\pi q\epsilon,
 \delta B_i=-\partial
_i \epsilon, \eqno(10)$$  where $q$ is an arbitrary constant.
\footnote
 {\dagger}{A simmilar translational gauge was used in $\lbrack 2,3
 \rbrack$}
 The full action (in addition to the determinant in (8)) is now:
$$S=L_{0,k_1}(h_1)+L_{g.f.}(\theta
,\phi)+{k_1\over 2\pi}\int d^2\sigma
 B_+(b\partial_-a-a\partial_-b-u\partial _-v+v\partial_-u
 $$$$
 +2{k_2\over k_1}q \cos^2\theta  \partial_-\phi)
 +B_-(b\partial_+a-a\partial_+b+u\partial _+v-
 v\partial_+u
 +2{k_2\over k_1}q
 \cos^2\theta \partial_+\phi) $$$$
 +B_+B_-(4-4uv+4{k_2\over k_1}q^2
 \cos^2\theta)\eqno(11a)$$
with $$L_0(h_1)=
-{k\over 4\pi}\int d^2\sigma\{\partial_+u\partial_-v
+\partial_-u\partial _+v
+\partial_+a\partial _-b
+\partial_-a\partial_+ b $$
$$-2\log u(
\partial_+a\partial _-b -\partial_-a\partial_+ b)\}
\eqno(11b)$$
Following Witten $\lbrack 1\rbrack$, according to the sign of $1-uv$
we can fix a gauge $a=\pm        b$. After integrating the gauge
fields out, we obtain the following action:   $$
 I=\int D[u,v,\theta,\phi]
 e^{S_{BH}}\det \lbrack {\pi^2\over 8k_1(
  1-uv+ Q^2\cos^2\theta)}\cdot
   {-\pi^2\over k_2\cos^2\theta}
  \rbrack  $$ where
$$S_{BH}(u,v,\theta,\phi) ={-k_1\over 8\pi}
\int d^2\sigma\{{Q^2\cos^2\theta(v^2\partial_+u
\partial _-u+ u^2\partial_+v\partial _-v)
\over (1-uv)(1-uv+Q^2\cos^2\theta)} $$
$$+{(2(1-uv)+Q^2\cos^2\theta(2-uv))
(\partial_+u\partial_-v+\partial_-u\partial_+v)
\over (1-uv)(1-uv+Q^2\cos^2\theta)} $$
$$+{k_2\over 2\pi}\int d^2\sigma
{(1-uv)\cos^2\theta\over 1-uv+Q^2\cos^2\theta}
\partial_-\phi\partial_+\phi
+\partial_-\theta\partial_+\theta$$
$$ + 2Q\sqrt{k_1\over k_2}
\cos^2\theta {  \partial_-\phi(u\partial_+v
-v\partial_+u)-
\partial_+\phi(u\partial_-v
-v\partial_-u)\over
1-uv+Q^2\cos^2\theta}\} \eqno(12)$$
with $Q^2={k_2\over k_1}q^2$.
We can calculate explicitly the determinant [7],
using the heat kernel
method.  To one loop order
 we obtain $$\det\lbrack {-\pi^4\over
4k_1k_2(1-uv+Q^2\cos^2\theta)\cos^2\theta}\rbrack
$$ $$
=\exp\{ -{1\over 8\pi}\int
d^2\sigma\sqrt h R^{(2)}\log(\cos^2\theta(
 1-uv+Q^2\cos^2\theta))+const.\}
\eqno(13)$$ where $R^{(2)} $ is the curvature
of the worldsheet and $h$ is the determinant of
the metric of the
worldsheet. The contribution of the Dilaton arises from (13).

After  redefining the fields:
 $$u=e^{t\sqrt{k_1\over k_2}
 }\sqrt{r-1},\hfill \;\;\;\;\;
 v=-e^{-t\sqrt{k_1\over k_2}}\sqrt{r-1}\eqno(14)$$
 the action becomes:
$$S_{BH}={k_2\over 2\pi}\int d^2\sigma
\{-{(r-1)(1+Q^2\cos^2\theta)\over r+Q^2\cos^2\theta^2}
\partial_- t\partial_+ t
 +{{k_1\over 4k_2}
 \partial_-r\partial_+r\over r(r-1)}$$ $$
 + {r\cos^2\theta\over r+Q^2\cos^2\theta}\partial_-\phi\partial_
+\phi +\partial_-\theta
\partial_+\theta $$ $$
+Q {(r-1)\cos^2\theta\over r+Q^2\cos^2\theta}
(\partial_-\phi\partial_+ t -
\partial_+\phi\partial_-t) \}\eqno(15)$$
 We can now shift  $\theta\rightarrow \theta+\pi/2$.
 Hence we have obtained the following
  space time metric  $d^2s={k_2\over 2}d^2\sigma$
where
$$d^2\sigma=-{(r-1))(1+Q^2\sin^2\theta)
\over r+Q^2\sin^2\theta}d^2t
+{{k_1\over 4 k_2}
d^2r\over r(r-1)}
+{r\sin^2\theta\over r+Q^2sin^2\theta}
d^2\phi +d^2\theta \eqno(16)$$
 the antisymmetric tensor (the "axion field")
 which has only the $t,\phi$ component
 $$B_{t\phi}=2k_2Q
 {(r-1)\sin^2\theta\over r+Q^2\sin^2\theta}\eqno(17)$$
 and the Dilaton $$\Phi=\log((r+\sin^2\theta)\sin^2\theta) +a
 \eqno(18)$$  where $a$
is an arbitrary constant.
 The space time metric (16) describes an axisymmetric black hole.
 $r=1$ is the event horizon and $r+Q^2\sin^2\theta=0$ is
 the  singularity. A simmilar type of
 singularity occures in the Kerr solution, where the singularity there
 is at $r+a\cos ^2\theta=0$, with $a$ being the total angular
 momentum devided by the mass.
 It can be seen there that
 the singularity is  a ring ("ring singularity").
  In our solution the black hole does not
  posses angular momentum, and in fact,  we shall see that $Q$ is related
  to the axion charge.
  We can define  a new coordinate
 $R=r+Q^2\sin^2\theta$, which is like the Boyer Lindquist
coordinates for the Kerr solution  $\lbrack 8\rbrack$.
  Then $R=0$ is the singularity and
 the event horizon is at $R=Q^2\sin^2\theta+1$.
 The above observations are obtained as a result of the fact that
 all the  scalar curvatures of the metric (Riemann curvature, Ricci
 curvature, scalar curvature)
   are     singular only at the point $R=r+\sin^2\theta=0$.
  The expressions are rather long, thus we bring here only
  the scalar carvature
   $$R_{curvature}=R_{\alpha}^{\;
   \alpha}={\hat R_1(\theta)\over \Delta^2}
   +{\hat R_2(\theta)
   \over \Delta^2 R}+{\hat R_3(\theta)\over \Delta^2R^2}+
   {24Q^2-12-9/a\over R^2}
   \eqno(20)$$ where
   $$R=r+Q^2\sin^2\theta$$
   $$\Delta=1+Q^2\sin\theta$$
   $$\hat R_1(\theta)= 4Q^4\sin^4\theta+8Q^2\sin^2\theta-2(Q^2-1)$$
   $$\hat
   R_2(\theta)=-Q^6sin^6\theta(16a-1)+Q^4\sin^4\theta (12Q^2a-28a+5)
   +17 aQ^2\sin^2\theta (Q^2-1)$$
  $$\hat R_3(\theta)=
  Q^2\sin^2\theta (Q^6\sin^6\theta(12a-3)+Q^4\sin^4
   \theta(2a-12Q^2a-9)+12Q^2a-4a+15)$$
  and $a=k_1/4k_2$.

The metric is not assymptotically flat. At $r\rightarrow \infty
$ the metric approaches
$$d^2S=-(1+Q^2\sin^2\theta)d^2t+{1\over r^2}(d^2 r+r^2(\sin^2\theta
 d^2\phi+d^2\theta))\eqno(21)$$
  Therefore, there is a distribution
of matter all over the space-time.

Now, in terms of the new ccordinate $\hat r=r+Q^2\sin^2\theta$ the
metric is:
$$d^2S=-{(\hat r-Q^2\sin^2\theta -1)(1+Q^2\sin^
\theta)\over \hat r}d^2t+
{ad^2\hat r
-aQ^2\sin 2\theta d\hat r d\theta\over
 (\hat r-Q^2\sin^2\theta)(\hat r-Q^2\sin^2\theta-1)}$$$$
+{\hat r
^2-\hat r(1+Q^2\sin^2\theta)
 +Q^2\sin^2\theta (4aQ^2-(4a-1)Q^2\sin^2\theta+1)\over
 (\hat r-Q^2\sin^2\theta)(\hat r-Q^2\sin^2\theta-1)}d^2\theta$$$$+
{(\hat r-Q^2\sin^2\theta)\sin^2\theta\over \hat r}d^2\phi\eqno(22)$$
($\hat r=Q^2\sin^2\theta$ is just a coordinate singularity).

We can see that $Q$ is related to
the axionic charge. Like the electromagnetic charge, we can define two
axionic charges (analogous to electric charge and magnetic charge).
One conserved charge that results from the effective  action is
 (in form notations)
$$\tilde q_{axion}=\int  e^{\Phi}{^*H} \eqno(24a)$$ where $^*H$ is
$H$ dual,  and $H=dB$. In four dimensions  $H$-dual is a one form
and the integral is around a closed curve. Therefore, this charge
vanishes, unless there is a string-type singularity. In our case
there is a point singularity and therefore this charge vanishes.
(For example, in three dimensions this becomes a constant (see
$\lbrack 3\rbrack$).)

A second conserved charge is
$$q_{axion}=\int Hd^3S \eqno(24b)$$
In our case this gives $q_{axion}=(16k_2\pi/3)Q$. Thus, the constant
$Q$ in the metric is proportional to the axion charge.

 The temperature of the black hole can be obtained by calculating
 the surface gravity [9]. The metric is static and the Killing
 vector
 $\partial _t^a$, which we denote $\chi^a$, defines a quantity
 $\kappa$ that is constant on the horizon. This quantity is
  associated with the temperature of the black hole.
   The surface gravity is defined as
   $$\kappa ^2=\lim \{-(\chi^b\nabla_b\chi^c )(\chi^a\nabla_a
  \chi _c)/\chi^d \chi_d\}\eqno(24)$$
   where $\lim$ stands for the limit as one approaches the
  horizon. In our case we obtain $$\kappa=T=
   \sqrt{{2\over k_1}}
 \eqno(25) $$
  For comparizon, the temperature of the two dimensional
 black hole [10] is $\sqrt {2\over k}$.

  Before we proceed, we want to mention that the conformal theory
  that describe our  solution (8) has a central charge
  $$c={3 k_1\over k_1-2}+{3k_2\over k_2+2}-2,\;\;\;k_2\in Z
  \eqno(26)$$
 (as     the central charge of $SL(2,R)$ is ${3k\over
k-2}$, the central charge of $SU(2)$ is ${3k\over k+2}$ and the
two gauged $U(1)$  decrease the central charge by 1 each).
   Hence we have a $k_1$ for every integer $k_2$ so that $c=26$.

 {\bf 3.  Exact Solution of 4D Black Membrane}

 Now we want to derive another solution of a four dimensional space
time metric.  In the following we start with a WZW model of the
group $SL(2,R)\otimes U(1)^2$.
 In order to obtain a Lorenzian signature we
  parametrise the $SL(2,R)$ group as before
by $$g_{SL(2,R)}=\left(
\matrix {a&u\cr -v&b\cr}\right)\eqno(27)$$
and the two $U(1)$ fields by $x$ and $y$.
The ungauged action is
$$L_0(g,x,y)=-{k\over 4\pi}\int d^2\sigma\{\partial_+u\partial_-v
+\partial_-u\partial _+v
+\partial_+a\partial _-b
+\partial_-a\partial_+ b $$
$$-2\log u(
\partial_+a\partial _-b -\partial_-a\partial_+ b)\}
+{k\over 2\pi} \int d^2 \sigma(
\partial_+x\partial _-x +
\partial_+y\partial _-y)   \eqno(28)$$
 We now gauge the $U(1)$ subgroup of $SL(2,R)$ together with
 a rotational symmetry in $x$ and $y$, i.e. we gauge the
 folowing symmetry transformation:
$$\delta a=2\epsilon a, \delta b =-2\epsilon b,
\delta u=\delta v=0,
 \delta x=-2q\epsilon y,\delta y=2q\epsilon x ,
 \delta A_i=-\partial
_i \epsilon, \eqno(29)$$  where $q$ is an arbitrary constant.
 The full action is now:
$$S=L_0(g,x,y)+{k\over 2\pi}\int d^2\sigma
 A_+(b\partial_-a-a\partial_-b-u\partial _-v+v\partial_-u
 +2q(x\partial_-y-y\partial_-x))$$
 $$+A_-(b\partial_+a-a\partial_+b+u\partial _+v-
 v\partial_+u
 +2q(x\partial_+y-y\partial_+x))
 +A_+A_-(4-4uv+4q^2(x^2+y^2))\eqno(30)$$
As in (12), according to the sign of $1-uv$
we can fix a gauge $a=\pm        b$. After integrating the gauge
fields out, we obtain the following action:   $$
 I=\int D[u,v,x,y] e^{S_{BM}}\det \lbrack {-\pi^2\over 8k(
  1-uv+ q^2(x^2+y^2)} \rbrack  $$ where
$$S_{BM}(u,v,x,y) ={-k\over 8\pi}
\int d^2\sigma\{{q^2(x^2+y^2)(v^2\partial_+u
\partial _-u+ u^2\partial_+v\partial _-v)
\over (1-uv)(1-uv+q^2(x^2+y^2))} $$
$$+{(2(1-uv)+q^2(x^2+y^2)(2-uv))
(\partial_+u\partial_-v+\partial_-u\partial_+v)
\over (1-uv)(1-uv+q^2(x^2+y^2))} $$
$$-4{(1-uv+q^2x^2)\partial_-x\partial_+x
+(1-uv+q^2y^2)\partial _-y
\partial_+y -q^2xy(\partial_-x\partial_+y
+\partial_+x\partial_-y)
\over 1-uv+q^2(x^2+y^2)}$$
$$ + 2q
{(x\partial_- y-y\partial_-x)(u\partial_+v
-v\partial_+u)-
(x\partial_+y -y\partial_+x)(u\partial_-v
-v\partial_-u)\over
1-uv+q^2(x^2+y^2)}\} \eqno(31)$$
The determinant to one  loop oredr is obtained as in (13)
$$\det\lbrack {-\pi^2\over
8k(1-uv+q^2(x^2+y^2)}\rbrack=$$$$=
\exp\{ -{1\over 8\pi}\int
d^2\sigma\sqrt h R^{(2)}\log( 1-uv+q^2(x^2+y^2)) +const.\}
\eqno(32)$$ which introduces
the contribution of the Dilaton field to the action.
Now we redefine the fields:
 $$u=e^{t/2}\sqrt{r-1},\;\;\;\;v=-e^{-t/2}\sqrt{r-1} $$
$$x={\rho\over 2}\sin \phi  ,\;\;\; y={\rho\over 2}\cos \phi
,\;\;\; \phi\in [0,2\pi] \eqno(33)$$
After defining the constant $Q=q/2$
 the action becomes:
$$S_{BM}={k\over 8\pi}\int d^2\sigma
\{-{(r-1)(1+Q^2\rho^2)\over r+Q^2\rho^2}
\partial_- t\partial_+ t
 +{\partial_-r\partial_+r\over r(r-1)}+
{r\rho^2\over r+Q^2\rho ^2}\partial_-\phi\partial_
+\phi $$ $$
+\partial_-\rho
\partial_+\rho
+Q {(r-1)\rho^2\over r+Q^2\rho^2}(\partial_-\phi\partial_+ t -
\partial_+\phi\partial_-t) \}\eqno(34)$$
 Hence we have obtained the following
  space time metric  $d^2s={k\over 8}d^2\sigma$
where
$$d^2\sigma=-{(r-1)(1+Q^2\rho^2)\over r+Q^2\rho^2}d^2t
+{d^2r\over r(r-1)}
+{r\rho^2\over r+Q^2\rho^2}d^2\phi +d^2\rho \eqno(35)$$
 the antisymmetric tensor which has only the $t,\phi$ component
 $$B_{t,\phi}={k\over 8}
 Q{(r-1)\rho^2\over r+Q^2\rho^2}\eqno(36)$$
 and the Dilaton $$\Phi=\log(r+Q^2\rho^2) +a\eqno(37)$$  where $a$
is an arbitrary constant.
 The metric is ill behaved at $r=1$,  which turns out to be the  event
 horizon, at $r=0$, or $\rho=0$ which turn  out to
  be merely   coordinate singularities and a true singularity
 at the point $(r,\rho)=(0,0)$. This is seen by calculation
 all the  scalar curvatures of the metric,
  which are all singular only at the point $(r,\rho)=(0,0)$.
 Again, we bring here only the scalar curvature
   $$R=R_{\alpha}^{\;\alpha}={
   2r-{5\over 4} +2Q^4\rho^2-{3\over 2}Q^2\rho^2+10Q^2
   \over r+Q^2\rho^2} +
   {Q^2\rho^2({3\over 2}r-{5\over 4}-14Q^2)\over
  (r+Q^2\rho^2)^2}$$
  $$+{2Q^2\rho^2(r-1)\over (1+Q^2\rho^2)^2}-2Q^2-{2Q^2\over
  1+Q^2\rho^2} + {9Q^4\rho^2-Q^2r\over
  (r+Q^2\rho^2)(1+Q^2\rho^2)}
  \eqno(39)$$
 The area
of the horizon is $\int\sqrt{g_{\phi\phi}g_{\rho\rho}}d\phi d\rho$
at $r=1$.
 Therefore,
unlike the first solution we found (12), the area of the horizon
in this space time is infinite, and therefore the horizon is
an infinite sheet surrounding the point of singularity.
Hence we have
a solution that describes a black membrane [11].

In this theory the central charge is $$c={3\ k\over k-2}-1+2=
4+{6\over k-2} \eqno(40)$$
 Hence for $k={25\over 11}$ $c=26$.

To see the asymptotic behavior of this metric we can define $z=
\log r$. Then for $z\rightarrow \infty$ we have
$$d^2S=-(1+Q^2(x^2+y^2))d^2 t+d^2z+d^2x+d^2y  \eqno(41)$$
The temperature of this black membrane is $T=\sqrt{2\over k}$, the same
 temperature as this of the 2D black hole.

{\bf 4.
Derivation Of a Solution of 4D Black Hole with Electromagnetic Field}

We turn now to describe a black hole solution with an
electromagnetic field.  Here we use the method used by Ishibashi
  in [2]. We can describe closed strings which have
 gauge fields in their massless spectrum in the following way
 $\lbrack 12\rbrack$.
  The action
contains the fields $X^{\mu}$ which describe  the space time coordinates
 and compactified free boson fields $Z^I$ which realisez the Kac-Moody
currents of the gauge group. The sigma model action is
$$S={1\over 2\pi} \int d^2\sigma g_{\mu\nu} \partial _+X^{\mu}
\partial_-X^{\nu} +\partial_-Z^I\partial_+ Z^I $$ $$+
A^I_{\mu}\partial
_+X^{\mu}\partial_ -Z^I +\tilde A^I_{\mu}\partial _-X^{\mu}\partial
_+Z^I   \eqno(42)$$
We are going to derive
a black hole in Lorenzian  signature.
The action contains the compactified  field $Z$,  related
to a $U(1)$ symmetry, which we eventually identify with
the electromagnetic
 field. The free lagrangian we start with now is
 $$L_0= WZW(SL(2,R)\otimes SU(2))+
 {1\over 2\pi}\int d^2\sigma
 \partial_+Z\partial_-Z\eqno(43)$$ where $Z$ is
compactified, namely $Z\sim Z+2\pi c$.  As before, we  gauge first
 a  $U(1)$ subgroup of $SU(2)$,
 which is the same gauge transformation as in
(6), and then we gauge a $U(1)$ subgroup of $SL(2,R)$
together with  a translational gauge in $\phi$ and
in $Z$, namely $\delta Z=
2 c\epsilon$\ together with (10).
The gauged lagrangian is (we write $k$ in place of $k_1$)
$$S=L_0+{k\over 2\pi}\int d^2\sigma
 B_+(b\partial_-a-a\partial_-b-u\partial _-v+v\partial_-u
 +2{k_2\over k}q\cos^2\partial_-\phi+{2c\over k}\partial_-Z)$$
 $$+B_-(b\partial_+a-a\partial_+b+u\partial _+v-
 v\partial_+u
 +2{k_2\over k}q\cos^2\theta\partial_+\phi+{2c\over k}\partial_+Z)$$ $$
 +4B_+B_-(1-uv+{k_2\over k}q^2\cos^2\theta+{c^2\over k})\eqno(44)$$
 As in (12), we fix a gauge $a=\pm b$.
 After integrating the gauge field out and defining $e^2={c^2\over
 k}$, $Q^2={k_2\over k}q^2$, we obtain
 $$S(u,v,\theta,\phi,Z)=I(u,v,\theta,\phi)$$$$
 + {e\over \pi}\int d^2\sigma
 \partial_+Z({2\sqrt{k_2}Q\cos^2\theta\partial_-\phi) - \sqrt{ k}
 (u\partial_-v-v\partial_
  -u)\over 1-uv+Q^2\cos^2\theta+e^2})$$ $$+
 \partial_-Z(2{\sqrt{k_2}
 Q\cos^2\theta\partial_+\phi +\sqrt {k}(u\partial_+v-v\partial_
  +u)\over 1-uv+Q^2\cos^2\theta+e^2})$$$$
  +{1\over 2\pi}\int {1-uv+Q^2\cos^2\theta\over
1-uv+Q^2\cos^2\theta +e^2}\partial_+Z\partial_-Z\eqno(45)$$
 where $$I(u,v,\theta,\phi)
 ={-k\over 8\pi}
\int d^2\sigma\{{(e^2+Q^2\cos^2\theta)(v^2\partial_+u
\partial _-u+ u^2\partial_+v\partial _-v)
\over (1-uv)(1-uv+e^2+Q^2\cos^2\theta)} $$
$$+{(2(1-uv)+(e^2+Q^2\cos^2\theta)(2-uv))
(\partial_+u\partial_-v+\partial_-u\partial_+v)
\over (1-uv)(1-uv+e^2+Q^2\cos^2\theta)} $$
$$+{k_2\over 2\pi}\int d^2\sigma
{(e^2+1-uv)\cos^2\theta\over 1-uv+e^2+Q^2\cos^2\theta}
\partial_-\phi\partial_+\phi
+\partial_-\theta\partial_+\theta$$
$$ + 2\sqrt{k\over k_2}Q
\cos^2\theta {  \partial_-\phi(u\partial_+v
-v\partial_+u)-
\partial_+\phi(u\partial_-v
-v\partial_-u)\over
1-uv+e^2+Q^2\cos^2\theta}\} \eqno(46)$$
and the Dilaton is $$\phi=\log((1-uv+Q^2\cos^2\theta+e^2)\cos^2\theta)
+a\eqno(47)$$ Now we redefine
 $$u=e^{t\sqrt{k/ k_2}}
 \sqrt {r-1-e^2} ,\;\;\;\; v=-e^{-t\sqrt{k/k_2}}
 \sqrt{r-1-e^2} $$
 and $\theta\rightarrow \theta+\pi/2$.
 Hence, the action describes a metric $d^2S={k_2\over 2}d^2\sigma$ with
  $$d^2\sigma=-{(r-1-e^2)(
  1+e^2+Q^2\sin^2\theta)\over r+Q^2\sin^2\theta}d^2t
+(1-{1+e^2\over r})^{-1} {kd^2r\over 4k_2r^2}$$ $$
+ {r\sin^2\theta
\over r+Q^2\sin^2\theta}
d^2\phi +d^2\theta,\eqno(48)$$  and
  $$B_{t \phi}=2k_2Q{(r-1-e^2)\sin^2\theta\over r
  +Q^2\sin^2\theta},\eqno(49a)$$
  the electromagnetic field
 $$A_{\phi}=2\sqrt{k_2}eQ{\sin^2\theta\over r+Q^2\cos^2\theta
 }\eqno(49b)$$
 $$A_t=2{k\over \sqrt{k_2}}e
 {(r-1-e^2)\over r+Q^2\sin^2\theta
 }\eqno(49c)$$ ($\tilde A_t=-A_t$ ; $\tilde A_{\phi}=A_{\phi}$),
  the Dilaton $\phi=\log((r+Q^2\sin^2\theta)\sin^2\theta)+a$, and
additional scalar field associated with the following vertex operator
$$\partial Z\bar \partial Z e^{i(-p_t t+p_r r+p_{\theta}
\theta+p_{\phi}
\phi)}.\eqno(50)$$

This describes a  black hole (the singularity is at $r+
Q\sin ^2\theta=0$ as in (16))
  which carries both axionic and electromagnetic charges.
The properties of this metric are very simmilar to those of
the one obtained
in (16), except for the fact that the event horizon is now at
$r=1+e^2$. Here we have an explicit expression for the electromagnetic
vector field and we therefore have the electromagnetic tensor
$F_{\mu\nu}$, defined by $$F_{\mu\nu}=\nabla_{\mu}A_{\nu}-\nabla
_{\nu}A_{\mu}.$$ The
electric ($F_{0,\mu}$) and the magnetic ($F_{i,\mu}$) fields are:
$$E_r=2{k\over \sqrt {k_2}}e{Q^2\sin^2\theta+1+e^2\over
 (r+Q^2\sin^2\theta)^2}\eqno(51a)$$
$$E_{\theta}=2{k\over \sqrt{k_2}}e{r-1-e^2 \over
(r+Q^2\sin^2\theta)^2}\eqno(51b)$$
 $$B_r= 2\sqrt{k_2}eQ{r\sin 2\theta\over (r+Q^2\sin^2\theta)^2}
 \eqno(51c)$$
 $$B_{\theta}=2\sqrt{k_2}eQ{sin^2\theta\over (r+Q^2\sin^2\theta)^2}
           \eqno(51d)$$
The central charge of this theory    is $$c={3k\over k-2}
+{3k_2\over k_2+2}-1\eqno(52)$$

Now, the effective action of this theory $\lbrack 13\rbrack$
can be written as
$$S_{eff}=\int\sqrt g e^{\tilde \phi}(R+(\nabla\tilde \phi)^2-
{1\over 4}(\nabla \psi)^2-e^{-\psi}({1\over 4}F^2+{1\over 12}H^2))
\eqno(53)$$
 where $\psi$ is the additional scalar field whose vertex operator
 is given in (50), $\tilde \phi=\phi+{1\over 2}\psi$, $F=dA$ and $H=
dB$. The electric charge is
 $$q_E=\int e^{\phi}{^*F}d^2 S
 \eqno(54a)$$
 where the integral is over a 2-sphere at infinity. This gives
 $q_E=
4\pi{k\over k_2}e\exp (a)C$, where $$C=\int_0^{\pi}d\theta\sin^3\theta
{Q^2\sin^2\theta+1+e\over \sqrt{1+Q^2\sin^2\theta}}$$
The  magnetic charge vanishes
 $$q_M=\int Fd^2S=0\eqno(54b).$$
Two axionic charges are associated with the action (53).
The first one, which vanishes in our solution is
$$\tilde q _{ax}=\int F\wedge F=0\eqno(54c) $$(although in our solution
locally $F\wedge F\ne 0$) and the other
axionic charge is $$q_{ax}=\int H d^3S=(16k_2\pi/3)Q\eqno(54d)$$
Thus,  this black hole carries both electric
and axionic charges but has no magnetic charge.

 Clearly, the equations of motion are  invariant
under the duality transformation $F\rightarrow
{ ^*F}$.
 Hence, magnetically charged black hole solution
may also be obtained as an equivalent string theory.

As a last comment, we would mention that applying exactly
the  same procedure in the $SL(2,R)\times U(1)^3/U(1)$ model leads
    to a solution of a black membrane
which carries  electric ard axionic charges as well.

\ack
I would like to express my gratitude to S. Yankielowic for
encouregment and interesting  discussions. I would also like to thank
N. Marcus,
G. Horowits, B. Reznik and A. Casher for helpfull
comments.
\vskip 2cm
\endpage
{\bf  REFERENCES}

\Item{1.} E.~Witten, "On String Theory and Black Holes",
preprint IASSNS-HEP-91/12,March 1991;
S.~Elitzur, A. Forgee and E. Rabinivici, Nucl.Phys.B359  (1991)
 581.
\Item{2.} N.~Ishibashi, M. Li and A.~Steif, "Two Dimensional
 Charged Black Holes in String Theory", preprint UCSBTH-91-28
   (1991).
\Item{3.} J.H.~Horne and
 G.~Horowitz, "Exact Black String Solutions in Three Dimensions",
 preprint UCSBTH-91-39   (1991)
\Item{4.} D.~Garfinkle, G. Horowitz and A. Strominger,
Phys.Rev.{\bf  43},3140   (1991)
\Item{5.} J.~Wess and B. Zumino, Phys.Lett.{\bf B37} (1971);
E. Witten, Comm.Math.Phys.
92  (1984) 455.
\Item{6.} P.~Goddard, A. Kent and D. Olive, Phys.Lett.{\bf B152}
 (1985) 88; W. Nahm, Duke Math.J.,54  (1987) 579;
E. Guadagnini, M. Martellini and M. Minchev, Phys.Lett.{\bf  191}
 (1987) 69; K. Bardakci, E. Rabinovici, B. Saring, Nucl.Phys.
{\bf  B299} (1988) 157;
D. Altshuster, K.Bardakci, E. Rabinivici, Comm.Math. Phys.
{\bf  118}
241; K. Gawedzki, A. Kupiainen, Phys. Lett. {\bf  B215} (1988) 119;
Nucl. Phys.
{\bf  B320}  (1989) 625;
D. Karabali, Q. Park, H. Schnitzer and Y. Yang, Phys.Lett.
{\bf  B216}
  (1989) 307;  H. Schnitzer, Nucl. Phys. {\bf  324}   (1989) 412;
D. Karabali and H. Schnitzer, Nucl.Phys.{\bf  B329}   (1990) 649.
\Item{7.} E.~Kiritsis, "Duality in Gauged WZW Models",
Berkly preprint, LBL-30747, May 1991.
\item{8.} Boyer and Lindquist, J.Math.Phys.8, 256 (1967).
\Item{9.} R.M. Wald, "General Relativity", The University of
Chicago Press, 1984.
\Item{10.} A. Giveon, "Target Space Duality and Stringy Black Holes",
 preprint LBL-30671, April 1991.
\Item{11.} Gary Horowitz, private communication.
\Item{12.} D.J. Gross, J.A. Harvey, E. Martinec and J.Perry,
Nucl.Phys.Lett.{\bf 54}, 502 (1985).
\Item{13.} C.G. Callan, D. Friedan, E.J. Martinec and M.J. Perry,
Nucl. Phys.{\bf B262}, 593, (1985).
\bye